# The Impact of Discretization Method on the Detection of Six Types of Anomalies in Datasets

Ralph Foorthuis

UWV, La Guardiaweg 116, 1040 HG Amsterdam, The Netherlands
`ralph.foorthuis@uwv.nl`

**Abstract.** Anomaly detection is the process of identifying cases, or groups of cases, that are in some way unusual and do not fit the general patterns present in the dataset. Numerous algorithms use discretization of numerical data in their detection processes. This study investigates the effect of the discretization method on the unsupervised detection of each of the six anomaly types acknowledged in a recent typology of data anomalies. To this end, experiments are conducted with various datasets and SECODA, a general-purpose algorithm for unsupervised non-parametric anomaly detection in datasets with numerical and categorical attributes. This algorithm employs discretization of continuous attributes, exponentially increasing weights and discretization cut points, and a pruning heuristic to detect anomalies with an optimal number of iterations. The results demonstrate that standard SECODA can detect all six types, but that different discretization methods favor the discovery of certain anomaly types. The main findings also hold for other detection techniques using discretization.

## 1   Introduction

*Anomaly detection* (AD) is the process of identifying cases, or groups of cases, that are in some way unusual and do not fit the general patterns present in the dataset [1, 2, 3]. The detection of *anomalies*, which are often also referred to as outliers, deviants or novelties, is a major research topic in the overlapping disciplines of artificial intelligence [4, 5, 6], data mining [7, 8, 9] and statistics [10, 11, 12]. It is not merely of interest for academia, however, as it is also of significant value in industrial practice nowadays [13, 14, 36]. Anomaly detection can be used for discovering fraud, data quality issues, security threats, process and system failures, and deviating data points that hamper model training.

   Many techniques for detecting anomalies have been devised throughout the years. The field of statistics traditionally focused mainly on parametric methods for discovering univariate outliers in each attribute (variable) separately [cf. 1, 12, 15]. Distance- and density-based techniques were consequently developed, allowing for non-parametric multidimensional data mining [16, 17, 18]. Another group of methods comprises complex non-parametric models, such as one-class support vector ma-chines, ensembles and various subspace methods [19, 20, 21]. Other approaches employ reconstruction techniques or information-theoretic concepts such as entropy and Kolmogorov complexity [22, 23]. Some solutions focus on individual cases (data points) [e.g. 16, 17, 25], whereas others aim to detect groups or substructures [e.g. 8, 23]. Discretization of continuous (numerical) attributes is a technique that is used in many of the AD approaches, e.g. for improving accuracy and time performance of the algorithms [24, 25, 26, 27, 28].

SECODA is an algorithm for unsupervised non-parametric anomaly detection in datasets with continuous and categorical attributes [25, 29]. It bears similarities with, i.a., density-based AD solutions and ensembles. SECODA employs discretization of numerical attributes, exponentially increasing weights and discretization cut points, as well as a pruning heuristic to detect anomalies with an optimal number of iterations. Its rich form of discretization makes it well-suited for this paper's experimentation.

This study investigates the effect of the discretization method on the unsupervised detection of each of the six anomaly types acknowledged in a recent typology of data anomalies [3]. The results not only demonstrate that SECODA, using its standard settings, is able to detect all six anomaly types, but also that different discretization methods clearly favor the discovery of different anomaly types. Moreover, the main results, as summarized in Table 2, also hold for other techniques using discretization.

This paper proceeds as follows. Section 2 presents the necessary theoretical background. Section 3 discusses the experiments that have been conducted with several synthetic and real-world datasets. Section 4 is for conclusions.

## 2  Theoretical Foundations

This section presents a summary of the typology of anomalies, a brief overview of discretization theory, and an explanation of the SECODA algorithm.

### 2.1  Typology of Anomalies

The typology of data anomalies presented in [3] offers a theoretical and tangible *understanding* of the nature of different types of anomalies, assists researchers with systematically *evaluating* the functional capabilities of anomaly detection algorithms, and as a framework aids in *analyzing* the nature of data, patterns and anomalies. The typology uses two fundamental and data-oriented dimensions:

- *Types of Data*: The data types of the attributes that are involved in the anomalous character of a deviant case. These can be *continuous* (numerical, e.g. height or temperature), *categorical* (code- or class-based, e.g. color or blood type) or *mixed* (when both types are involved).
- *Cardinality of Relationship*: The way in which the various attributes relate to each other when describing anomalous behavior. When no relationship between the variables exists to which the anomalous character of the deviant case can be attributed, the relationship is said to be *univariate*. It follows that the analysis can assume independence between the attributes. On the other hand, when the deviant behavior of the anomaly lies in the relationships between its variables, i.e. in the combination of its attribute values, then the relationship is said to be *multivariate*. This means the variables need to be analyzed jointly, not separately, in order to account for the relationships between them.

|  |  | **Types of Data** | | |
|---|---|---|---|---|
|  |  | **Continuous attributes** | **Categorical attributes** | **Mixed attributes** |
| **Cardinality of Relationship** | **Univariate** | Type I<br>Extreme value anomaly | Type II<br>Rare class anomaly | Type III<br>Simple mixed data anomaly |
|  | **Multivariate** | Type IV<br>Multidimensional numerical anomaly | Type V<br>Multidimensional rare class anomaly | Type VI<br>Multidimensional mixed data anomaly |

**Fig. 1.** The typology of anomalies

These two dimensions naturally and objectively yield six basic types of anomalies. Although the typology can be used to describe aggregate anomalies (a group of cases that deviates), the focus in this study is on individual data points.

- *Type I - Extreme value anomaly*: A case with an extremely high, low or otherwise rare (e.g. isolated intermediate) value for one or several individual numerical attributes. This type of outlier is typically considered in traditional univariate statistics, e.g. by using a measure of central tendency plus or minus 3 times the standard deviation or the median absolute deviation. Examples of Type I anomalies are the *Ia* and *Ib* cases in Fig. 2.A (note: the reader might want to zoom in on a digital screen to see colors, patterns and data points in detail).
- *Type II - Rare class anomaly*: A case with an uncommon class value for one or several individual categorical variables. Such values can be few and far between or truly unique (i.e. occur only once). An example of a Type II anomaly is the *IIa* case in Fig. 2.B, which is the only square shape in the set.
- *Type III - Simple mixed data anomaly*: A case that is both a Type I and Type II anomaly, i.e. with at least one extreme value and one rare class. This anomaly type deviates with regard to multiple data types. This requires deviant values for at least two attributes, each anomalous in their own right. These can thus be analyzed separately. Analyzing the attributes jointly is not necessary because, like Type I and II anomalies, the case is not deviant in terms of a combination of values. An example of a Type III anomaly is the *IIIa* case in Fig. 2.B, a unique shape at an extreme numerical position.

- *Type IV - Multidimensional numerical anomaly*: A case that does not conform to the general patterns when the relationship between multiple continuous attributes is taken into account, but that does not have extreme or isolated values for any of the individual attributes that partake in this relationship. The anomalous nature of a case of this type lies in the deviant or rare combination of its continuous attribute values. Detection therefore requires several numerical attributes that are analyzed jointly. An example of a Type IV anomaly is the *IVa* case in Fig. 2.A.
- *Type V - Multidimensional rare class anomaly*: A case with a rare combination of class values. A minimum of two categorical attributes needs to be analyzed jointly to discover a multidimensional rare class anomaly. An example is this curious combination of values from three attributes used to describe dogs: 'MALE', 'PUPPY' and 'PREGNANT'. Another example is the *Va* case in Fig. 2.B, which is the only red circle in the set.
- *Type VI - Multidimensional mixed data anomaly*: A case with a deviant relationship between its continuous and categorical attributes. The anomalous case generally has a categorical value or a combination of categorical values that in itself is not rare in the dataset as a whole, but is only rare in its neighborhood (numerical area) or local pattern. As with Type IV and V anomalies, multiple attributes need to be jointly taken into account to identify them. In fact, multiple datatypes need to be used, as a Type VI anomaly per definition requires both numerical and categorical data. Examples of Type VI anomalies are the *VIa* cases in Fig. 4.A, seemingly misplaced green cases amongst an overwhelmingly red data cloud.

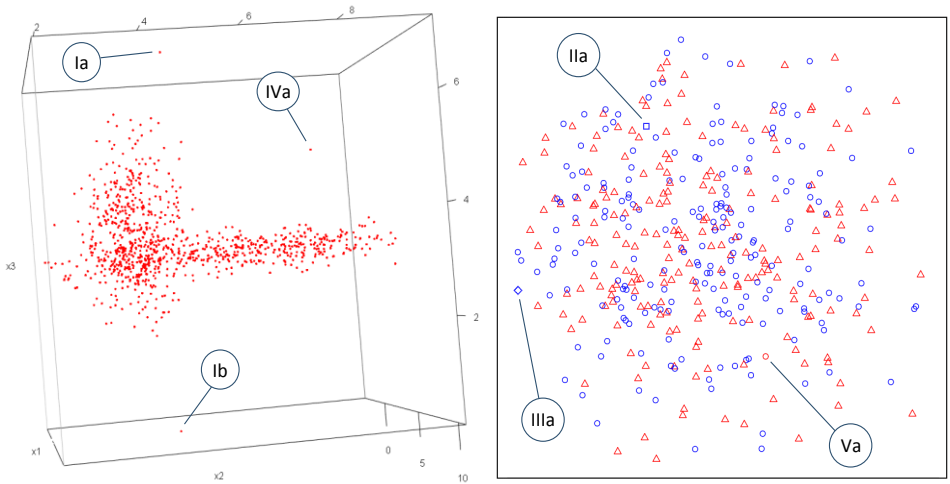

**Fig. 2.** (A) Mountain dataset with 3 numerical attributes; (B) ClassCircle dataset with two numerical attributes and two categorical attributes (color and shape)

The value of this typology lies not only in providing both a theoretical and tangible understanding of the types of anomalies one can encounter in datasets, but also in its ability to evaluate which type of anomalies can be detected by a given algorithm – or a given configuration of an algorithm. See [3, 25] for more examples of anomalies.

### 2.2 Discretization

The task of *discretization* refers to partitioning a continuous attribute into a limited number of sub-ranges (intervals) in order to obtain a categorical data type [27, 28, 30]. Discretization is used regularly in artificial intelligence, as numerous machine learning and data mining algorithms require a categorical feature space [7, 27, 28, 30]. Examples of algorithms where discretization plays a crucial role are decision trees, random forests, Bayesian networks, naive Bayes and rule-learners. Discretization also plays an important role in anomaly detection [cf. 24, 25, 26]. Apart from the fact that techniques may require categorical data, discretization has been shown to improve the accuracy, time performance and understandability of analysis methods [27, 28, 30].

The term *arity* refers to the resulting number of intervals or partitions. Several methods allow to set this number $b$ before running the discretization process. The range of a continuous variable is divided into intervals by $b - 1$ cut points. An individual *cut point* or split point is a real value at the position where an interval boundary is located, dividing the range into two intervals.

Discretization methods can be supervised, taking into account the training set's class label that ultimately needs to be predicted, or unsupervised, thus not taking into account a dependent variable. Two main unsupervised discretization methods exist, both of them often referred to as *binning* [7, 26, 27, 31]. *Equiwidth* discretization refers to equal interval binning. This method divides the range of an attribute's observed continuous values into $b$ bins of the same value interval. The second method is *equidepth* discretization, which refers to equal frequency binning and divides a continuous attribute into $b$ bins that each contain the same number of cases. In both methods $b$ is provided as input to the discretization function. The two discretization techniques have been used for anomaly detection [e.g. 24, 25, 26].

Discretization methods can be characterized in several ways [28, 30, 31]. Binning techniques can be global or local, albeit both unsupervised methods employed in this study are global. This means that they use the entire value space for partitioning, independently of the characteristics of local regions. Methods can also be direct or incremental, with the latter referring to techniques that pass through the data several times to arrive at an optimal discretized attribute. The equiwidth and equidepth methods are direct, meaning that they require only one pass. Finally, both binning methods discretize the data for each attribute separately, so these binning solutions do not take into account any relationships between the variables.

### 2.3 SECODA

SECODA, an abbreviation for segmentation- and combination-based detection of anomalies, is a general-purpose algorithm for unsupervised anomaly detection in datasets with mixed data [25, 29]. The algorithm is non-parametric in nature and there-fore does not assume any specific data distribution. It investigates the joint density distribution to discover high-density patterns and low-frequency deviations in the dataset, taking into account any relationship that may exist between its attributes. To this end, SECODA iteratively searches the dataset until the cases have been scrutinized with sufficient detail.

SECODA is guaranteed to identify cases with unique or very rare combinations of attribute values. The algorithm uses the histogram-based approach to assess the density of each combination (or "constellation") of categorical and continuous attribute values. The concatenation trick, which combines categorical and discretized continuous attributes into a new constellation feature, is used to analyze different data types in a joint fashion. In conjunction with recursive binning this captures complex relationships between attributes. In subsequent iterations SECODA uses increasingly narrow discretization intervals in order to add more detail and precision to the analysis and identify more subtle anomalies. The distance between data points in numerical space is implicitly accounted for by this iterative binning process. A pruning heuristic as well as exponentially increasing weights and arity are employed to speed up the analysis. The increasing arity (providing more localized details) and weights (allowing for optimally combining the results obtained from different iterations) also help to avoid discretization error and detection bias.

Note that recursive discretization is not employed by SECODA to find a single, optimal value for the arity parameter $b$, because it exploits the information from all binning iterations. Put differently, SECODA is an algorithm that recursively collects and uses the information from a discretization method that is itself applied in each iteration in a direct (instead of incremental) manner. The input parameter $b$ is thus not provided by the user, but repeatedly by SECODA until a stopping criterion is reached.

The SECODA approach has several favorable properties. It is a relatively simple algorithm that does not require expensive point-to-point calculations. Only basic data operations are used, making it suitable for sets with large numbers of rows as well as for in-database analytics and machines with relatively little memory. The algorithm scales linearly with dataset size, and for extremely large sets a longer computation time is hardly required because additional iterations would not yield a meaningful gain in precision. The technique can also easily be implemented for parallel processing architectures. All kinds of relationships between attributes are taken into account, such as (non)linear associations, interactions, collinearity and relations between variables of different data types. Although SECODA is vulnerable to the curse of dimensionality, general techniques such as feature bagging and random projection can be applied to deal with this. Missing values are automatically handled as one would functionally desire in an AD context, with only very rare missing values being considered anomalous. Finally, the pruning heuristic is a self-regulating mechanism during runtime, dynamically deciding how many cases to discard. After converging the algorithm returns a score vector so that each case gets assigned a degree of anomalousness, with lower scores representing more deviant occurrences.

SECODA has been evaluated in an academic context and has been used in practice as well to discover anomalies in the Polis Administration, an official register maintaining masterdata regarding the salaries, social security benefits, pensions and in-come relationships of people working or living in the Netherlands [25, 37]. The evaluation involved applying the algorithm to various synthetic and real-world datasets. Using ROC and PRC curves, as well as AUC and partial AUC metrics, it was demonstrated that this AD solution is able to successfully detect a wide variety of anomaly types. It has also been shown that the algorithm has low memory requirements and scales linearly with dataset size. SECODA has not been tested on all six types of anomalies, as the full typology was published later. Section 3 will demonstrate that the algorithm is indeed able to detect all types, and is therefore well-suited for experiments studying the effects of discretization on the detection of these types.

SECODA can be downloaded for free as a package for the R environment (see Remarks). The implementation offers various options, such as the minimum and maximum amount of iterations, a pruning parameter, and the iteration after which the heuristics should start to run. These options generally have trivial consequences and are mainly intended to tweak the amount of analysis detail and running time, so the standard settings normally suffice. This is desirable because algorithms for data mining are ideally parameter-free in order to discover the true patterns and deviations in a simple and objective fashion [23, cf. 18]. On the other hand, however, it is widely acknowledged that the world – and therefore the datasets that it produces – is extremely complex, and that no single algorithm or algorithm setting is thus able to perform excellent in all situations [18, 32, 33, 34]. This also holds in the context of anomaly detection [35, 36] and discretization [30]. Section 3 therefore investigates the effect of the binning method, another parameter that the analyst can set before running SECODA, on detecting the different types of anomalies defined in section 2.1.

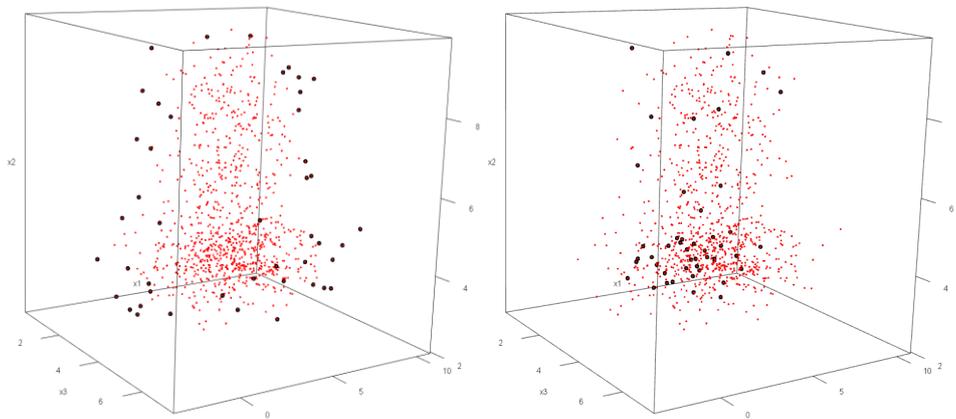

**Fig. 3.** (A) The large black dots represent the top 45 anomalies of the Mountain set resulting from equiwidth binning; (B) The top 45 anomalies from equidepth binning

## 3 Empirical Experiments

### 3.1 Research Design and Datasets

This study uses several synthetic and real-world datasets to investigate whether and how the discretization method affects the detection of the various anomaly types. The simulated datasets are labelled, which makes them suitable for verifying whether AD algorithms can readily detect the anomalies. The real-world dataset, drawn randomly from the aforementioned Polis Administration and anonymized subsequently, is unlabeled. The sets are described in Table 1 and are visually depicted in Figures 2 to 5. See the Remarks for download options. The R environment 3.4.3, RStudio 1.1.383, SECODA 0.5.3 and rgl 0.98.22 were used to generate the synthetic datasets and conduct the experiments. SECODA's heuristics for speeding up the analysis (e.g. pruning, which in a standard configuration starts being applied after 10 iterations) were not used in order to ensure maximum precision of the results.

**Table 1.** Datasets used for experiments

| Dataset | Nature | Datatypes | # Cases | Types of anomaly |
|---|---|---|---|---|
| ClassCircle | Simulated | 2 num, 2 categ | 422 | Type II, III, V |
| Mountain | Simulated | 3 numerical | 943 | Type I, IV |
| NoisyMix | Simulated | 3 num, 2 categ | 3867 | Type II, VI |
| Sword | Simulated | 2 num, 1 categ | 7024 | Type II, III, VI |
| Helix | Simulated | 3 num, 1 categ | 1410 | Type I, IV, VI |
| Polis dataset | Real-world | 3 num, 1 categ | 304726 | Type I, II, IV, VI |

Although the multivariate anomaly types can be used to describe aggregate anomalies – i.e. a group of related cases that deviates as a whole [3] – this study will focus solely on deviants that are atomic, single cases in independent data. The reason for this is that detecting grouped anomalies generally requires special-purpose approaches.

### 3.2 Results and Discussion

In the first series of experiments the five simulated datasets were used to study whether SECODA was able to identify the six types of anomalies presented in section 2.1. Note that [25] was not able to evaluate the algorithm on all six types because the full typology of [3] had not been developed at the time. The standard configuration of SECODA employs equiwidth binning and was indeed able to detect all types of anomalies. The subsequent series of experiments involved running SECODA with the non-standard equidepth setting to investigate what types of anomalies were identified in this fashion and how this compared to equiwidth AD.

With regard to a univariate analysis of a *single numerical attribute*, it is evident that the equiwidth setting is the preferred and basically only sensible option. This setting is able to detect isolated Type I cases, both extremely large or small values and rare intermediate data points. The equidepth setting, even though many discretization iterations were generally required before converging, was not able to detect these obvious anomalies and resulted in all cases getting a very high and non-discriminating score. This can be easily explained by the nature of equidepth discretization, since every bin gets assigned the same number of cases (although slight differences in frequency might occur if the set cannot be split evenly). SECODA's repeated binning with increasingly narrow intervals does not change this fact.

For the Mountain set with *multiple numerical attributes* the equiwidth setting was also found to be the superior choice, as it was readily able to detect the 3 labelled Type I and IV anomalies. Furthermore, the other cases with a low score were all relatively isolated cases at the boundary of the data cloud. With the equidepth setting only 1 of the 3 labelled anomalies were detected (the Type IV case of Fig. 2.A). In addition, most of the other low-score equidepth results were positioned in the middle of the data cloud, seemingly without a good reason why these should be considered more anomalous than other data points. The difference between the two binning methods is illustrated by Fig. 3.A on the left depicting the 45 most anomalous cases found by equiwidth binning, which are mostly outlying and include the 3 labelled anomalies, and 3.B showing the 45 lowest-score cases, which are mainly positioned in the high-density center of the data cloud. (Note that the aforementioned 3 true anomalies, which can be clearly seen in Fig. 2.A, are not visible from this angle.)

When disregarding the categorical attributes in the Helix and NoisyMix sets, the results are very similar. Type IV anomalies can be detected relatively well by equidepth binning, albeit with more false positives. Type I deviants are not detected, although they may be found if they have extreme values for multiple numerical attributes and thus are anomalous with regard to the combination of these values.

In short, the equidepth setting is most certainly not suitable for AD analysis of univariate numerical vectors (hosting Type I cases) and is reasonably equipped for dealing with multivariate numerical sets (hosting Type IV cases). Equiwidth binning yields more meaningful results as it directly targets the numerically isolated cases.

When analyzing a dataset containing *only categorical attributes*, the discretization method does not in any way influence the results. This is entirely to be expected, as discretization of continuous data should not affect a purely categorical analysis. The binning method provided by the analyst as an input parameter to the algorithm is simply irrelevant in this situation. Tests on several datasets indeed confirm this when running the algorithm with the two settings. In sets with *mixed data* both numerical and categorical attributes are present, and the returned scores of the two discretization methods can be expected to be different. However, the effect depends on the type of anomaly and the distribution of the data. Truly unique Type II or III univariate class-based anomalies will be recognized as unique, regardless of the binning method, and get assigned the lowest score possible. The same holds for unique combinations of classes, i.e. Type V cases. Experiments with the datasets that contain categorical data confirmed this as well, with SECODA returning the lowest anomaly score for such unique cases with both methods. However, when the Type II, III or V anomalies are rare in the dataset (rather than truly unique), the numerical data may influence the score. This can be expected because the rare cases can be close or distant neighbors and also compete with e.g. very isolated Type I and VI deviants. However, regardless of the binning method one would still expect these anomalies to be identified, returning relatively low anomaly scores for such cases. This is confirmed as well, although with some interesting differences between the two discretization methods (see below).

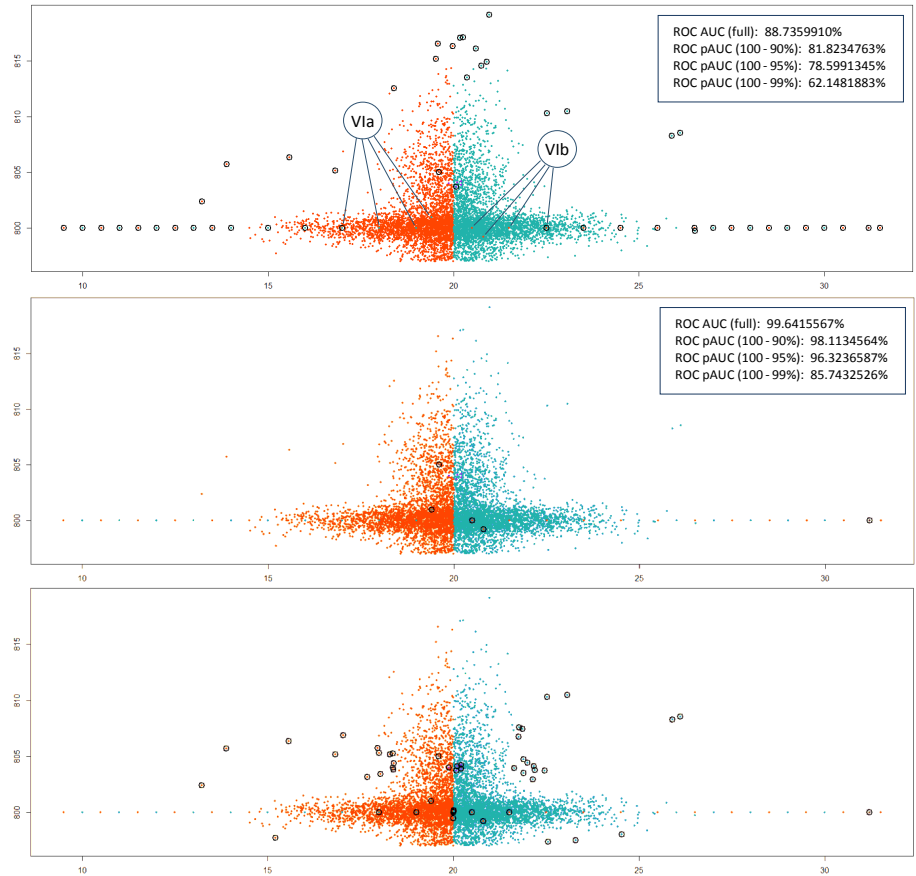

**Fig. 4.** From top to bottom: (A) The top 50 anomalies (black circles) of the Sword set from equiwidth binning; The (B) top 5 and (C) top 50 anomalies from equidepth binning

The binning method possibly has the most interesting impact on the detection of Type VI anomalies. These do not feature truly (globally) unique classes, because these classes are common in other areas of the numerical space. The detection of these local anomalies may therefore very well be affected by the discretization technique, an expectation that was confirmed by the experiments. In several datasets it was observed that equidepth binning often yields superior results when the goal is to detect Type VI anomalies. This is illustrated by Fig. 4.A at the top, where it can clearly be seen that the equiwidth analysis results in a variety of anomalies. However, due to the nature of the Sword dataset, which contains many numerically isolated cases, most of the top 50 anomalies are Type I and IV outliers. The Type II and III anomalies are also detected, but the Type VI anomalies less so. The equidepth analysis presented in Fig. 4.B and 4.C results in quite different cases being denoted as most extreme anomalies. It can be seen that the top 5 cases are mostly Type VI anomalies, which are located in dense (rather than sparsely populated) regions of the space. The Type II case at height 805 and the Type III case at the far right are truly unique classes due to their color and are therefore regarded as highly anomalous by both binning methods. Rare (as opposed to truly unique) Type II and V anomalies, which can but do not have to be isolated, are also detected more readily with equidepth binning when not located in low-density areas. Equiwidth binning will acknowledge a handful of neighboring rare classes (i.e. a very small 'cluster') as moderately anomalous, regardless of whether they lie inside or outside the data cloud. This is due to the fact that they are not truly unique. Equidepth binning, on the other hand, will recognize them as highly anomalous if they lie within the cloud, but not if they lie outside it (see the five detected purple cases in the middle of Fig. 4.C). Fig. 4 also shows the ROC AUC and 3 specificity partial AUCs for the specific task of detecting the in-cloud high-density anomalies (not the numerically isolated cases). In short, equiwidth discretization is well-equipped for detecting all anomaly types, including isolated occurrences. Equidepth binning, although more vulnerable to yielding false positives, is relatively well-equipped for detecting Type VI and in-cloud Type II and V anomalies.

To further investigate these findings, SECODA was used to analyze a sample from the aforementioned real-world Polis dataset. A similar effect was observed here. Fig. 5.A on the left illustrates the results of AD with equiwidth binning, which yielded a wide variety of anomalies, including many isolated cases. Fig 5.B shows the results of AD with equidepth discretization, with the most extreme anomalies found to be positioned in the center's high-density area. Both figures also have a zoomed-in view at the bottom, where the difference can be seen in more detail for each binning method.

At this point it is valuable to discuss the reasons why equiwidth and equidepth discretization yield different results in an AD context. In general, equiwidth binning performs better in terms of overall functional performance, i.e. the capability to detect a wide variety of meaningful anomalies. The reason for this is that equiwidth binning (or at least a single binning run with only one value for *b*) uses fixed value intervals, resulting in isolated Type I, III and IV cases to be placed in near empty bins. This also holds when SECODA repeatedly discretizes the continuous attributes using many values for *b* during the analysis, thus creating few bins in early iterations and many bins in later iterations (the recursive binning ensures that more distant anomalies get lower scores). It is known from the literature that data analysis with equiwidth binning is sensitive to outliers, a property that is usually seen as a disadvantage [7, 27, 28, 31]. However, in the context of anomaly detection this sensitivity can be exploited, resulting in relatively easy detection of sparse data by isolating them in separate bins. Equidepth binning, on the other hand, fails this detection of isolated cases, since the value ranges of the bins are stretched so as to fill them with an equal amount of data points. For example, in a typical Gaussian distribution the discretization intervals at the tails will be very wide because these regions are sparsely populated and the bins have to be filled with a given amount of cases. Moving inwards to the mean of the Gaussian distribution the bins will get narrower. The consequence is that univariately numerically isolated cases are not detected, as the focus is then on the categorical abnormality and − in case of multivariate analysis − on the combination of values from multiple attributes. Compared to equiwidth binning this results in (univariate) low-density cases getting assigned relatively high SECODA scores and non-isolated deviant cases relatively low scores.

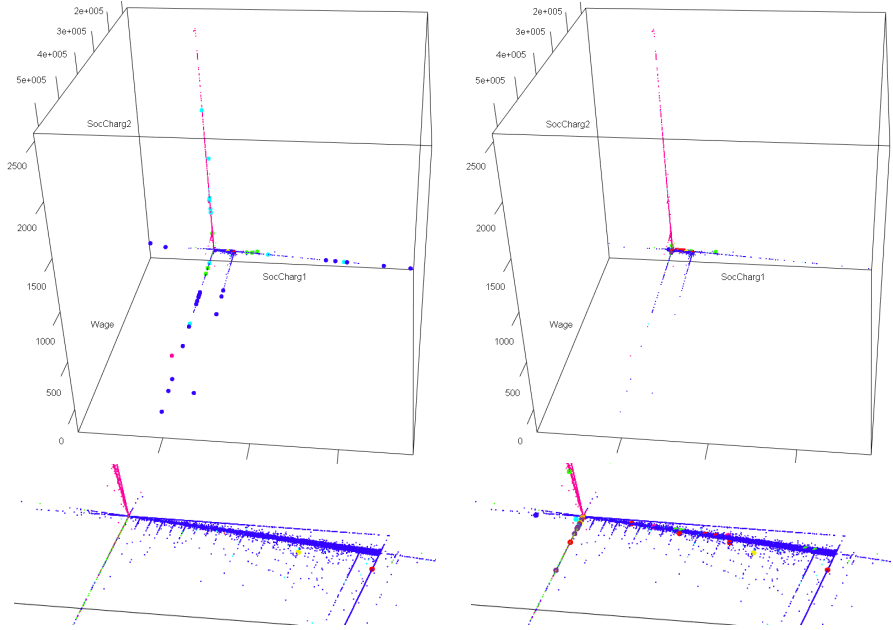

**Fig. 5.** (A) The large dots represent the 40 most extreme anomalies of the Polis set detected by equiwidth binning (the bottom is zoomed-in); (B) The top 40 anomalies from equidepth binning

Moreover, in the narrow intervals used for univariate high-density areas, the class values of Type II, V or VI cases will be quickly (i.e. with a relatively low value for *b*) unique in its bin, even if the case is not located very distantly from the cases with a similar color. In Fig. 4, for example, the red Type VI anomalies somewhat left from value 20 are not located very far from the large amount of normal red cases that can be seen from value 20 and up. With equidepth binning the discretization intervals in that region of the variable plotted on the horizontal axis will be very narrow, resulting in earlier separation and therefore detection of these anomalies than would be the case with equiwidth binning.

It was mentioned above that equidepth binning recognizes several neighboring rare classes (referred to as the very small 'cluster') as very anomalous if they lie within the rest of the data cloud, but not if they lie isolated. This can be explained by the same reason: within the dense parts of the data cloud the discretized value intervals are narrower, so rare classes are recognized with a lower arity *b* than with equiwidth discretization. This means they get detected earlier and are denoted as more anomalous.

Equidepth binning in an AD context thus scrutinizes the dense regions of the distribution in more detail (if these can be detected univariately). This method of discretization disregards tail and intermediate data points that are isolated in numerical space. Instead, when a multivariate analysis is conducted, the focus will be on *uncommon class values* and *rare combinations* of (continuous and categorical) attribute values. Equidepth discretization thus ignores univariately isolated cases and, more so than equiwidth analysis, has a propensity to detect anomalies that lie amongst other data points. It favors detecting cases that w.r.t. numerical attributes are located in univariate high-density regions. The discretization process, which handles individual attributes, will place cases that are located in univariate high-density regions in very thin univariate bins, i.e. in narrow value intervals. If these cases are also located in the univariate high-density ranges of other attributes, the multidimensional intersection will thus yield relatively low-density, sparsely populated constellations. In a purely numerical dataset this property will denote as anomalies both Type IV cases (true deviants) and points in or around the densest areas of the data cloud (often false positives, but sometimes interesting subtle deviants). This argument holds both for one-time discretization and the iterative binning of SECODA. For mixed data this works well for discovering Type VI anomalies, as well as for Type II and V cases located in high-density areas.

To succinctly state why equidepth binning can discriminate between normal and anomalous cases: It is not because *bins get scarcely filled with isolated points*, because all bins are filled with an equal amount of cases. Rather, it is because *categorical data with an unbalanced class distribution* is present or because the *combination of numerical and/or categorical values* yields infrequent occurrences. Equiwidth binning, on the other hand, utilizes all these three discriminating properties.

**Table 2.** Impact of discretization method on detection of anomaly types

| Type | Impact? | Useful? | Explanation |
|---|---|---|---|
| I | Y | N | ED cannot discriminate between the univariate numerical values and is intrinsically not equipped to detect this type. |
| II | N/Y | Y | ED is identical to EW when analyzing a single categorical attribute. It can be more useful than EW if the goal is to detect (non-unique) rare Type II anomalies in numerically high-density regions in an analysis of mixed data. |
| III | Y | Y | ED detects truly unique classes equally well as EW, but the latter shows slightly better performance with rare classes (because EW will exploit their isolated position better). |
| IV | Y | Y | ED detects many Type IV cases, but also yields more false positives and false negatives, and is thus not optimally equipped to detect this type. ED can sometimes detect more subtle Type IV cases at dense areas than EW can. |
| V | N/Y | Y | ED is identical to EW in a set with merely categorical data. It can be more useful than EW if the goal is to detect (non-unique) rare Type V anomalies in numerically high-density regions in an analysis of mixed data. |
| VI | Y | Y | ED tends to favor the detection of Type VI anomalies and can be more useful than EW if this is indeed the goal. |

Table 2 summarizes the findings for each anomaly type. The *Impact?* column refers to whether there exists a direct impact of using equidepth (ED) instead of equiwidth (EW) binning for the anomaly type. The *Useful?* column denotes whether equidepth binning can be useful in some situations for detecting the given type.

These main conclusions also hold for single, non-iterative binning operations, e.g. using only 7 intervals to discretize each continuous attribute. However, the recursive binning of SECODA accounts for the distance between data points and is thus able to calculate the degree of deviation. A single discretization run, on the other hand, requires the analyst to pick an arbitrary number of bins and cannot return information on the degree of anomalousness as a result of this rather crude form of binning.

As a final note, equidepth discretization can be useful in practical situations, as it is known that in some settings it is valuable to detect non-isolated and relatively subtle deviations rather than cases that are extreme and rare on all accounts [cf. 38, 39].

## 4    Conclusion

The results demonstrate that discretization, including its employment in the standard SECODA algorithm, can be used to detect all six types of anomalies. However, the equiwidth and equidepth discretization techniques yield notably different results and favor the discovery of certain anomaly types. Equiwidth and equidepth SECODA can therefore best be seen as two different algorithms. Equiwidth SECODA is a general-purpose algorithm, whereas the equidepth version is a special-purpose technique focusing on specific anomaly types. The main conclusions of Table 2 also hold for techniques that perform discretization only once, although the results hereof will be less precise and will not account for the distance between data points.

In general, if the analyst does not know beforehand in what type of anomaly he or she is interested, then equiwidth discretization is the preferred option. This will conduct a general-purpose anomaly detection and ensure that all anomaly types will be detected. If on the other hand the focus is on identifying anomalies that are not located in extreme or isolated regions of the numerical space, equidepth discretization should be used. The equidepth binning option favors the detection of Type VI anomalies as well as Type II and V cases that are found inside data clouds rather than in sparsely populated regions.

**Remarks.** A SECODA implementation, as well as various datasets and the code to analyze them in R can be downloaded from www.foorthuis.nl (see "SECODA resources for R").

## References

1. Barnett, V., Lewis, T.: *Outliers in Statistical Data.* Third Edition. Chichester: Wiley (1994).
2. Goldstein, M., Uchida, S.: *A Comparative Evaluation of Unsupervised Anomaly Detection Algorithms.* PLoS ONE, Vol. 11, No. 4 (2016).
3. Foorthuis, R.: *A Typology of Data Anomalies.* Proceedings of the 17th International Conference on Information Processing and Management of Uncertainty in Knowledge-Based

Systems (IPMU 2018), Cádiz, Spain; Springer CCIS 854 (2018). DOI: 10.1007/978-3-319-91476-3_3

4. Pang, G., Cao, L., Chin, L.: *Outlier Detection in Complex Categorical Data by Modelling the Feature Value Couplings*. Proceedings of the 25th International Joint Conference on Artificial Intelligence (2016).
5. Riahi, F., Schulte, O.: *Propositionalization for Unsupervised Outlier Detection in Multi-Relational Data*. Proceedings of the 29th International Florida Artificial Intelligence Research Society Conference (2016).
6. Hengst, F. den, Hoogendoorn, M.: *Detecting Interesting Outliers: Active Learning for Anomaly Detection*. Proceedings of the 28th Benelux Conference on Artificial Intelligence, Amsterdam, the Netherlands (2016).
7. Tan, P., Steinbach, M., Kumar, V.: *Introduction to Data Mining*. Boston: Addison-Wesley (2005).
8. Noble, C.C., Cook, D.J.: *Graph-Based Anomaly Detection*. Proceedings of the Ninth ACM SIGKDD International Conference on Knowledge Discovery and Data Mining (2003).
9. Schubert, E., Weiler, M., Zimek, A.: *Outlier Detection and Trend Detection: Two Sides of the Same Coin*. Proceedings of the 15th IEEE International Conference on Data Mining Workshops (2015).
10. Hubert, M., Rousseeuw, P., Segaert, P.: *Multivariate Functional Outlier Detection*. Statistical Methods & Applications, Vol. 24, No. 2, pp 177-202 (2015).
11. Ranshous, S., Shen, S., Koutra, D., Harenberg, S., Faloutsos, C., Samatova, N.F.: *Anomaly detection in dynamic networks: A survey*. WIREs Computational Statistics, Vol. 7, No. 3, pp. 223-247 (2015).
12. Fielding, J., Gilbert, N.: *Understanding Social Statistics*. London: Sage (2000).
13. Gartner: *Hype Cycle for Data Science and Machine Learning, 2017*. Gartner, Inc (2017).
14. Forrester: *The Forrester Wave: Security Analytics Platforms, Q1 2017*. Forrester Research, Inc (2017).
15. Leys, C., Ley, C., Klein, O., Bernard, P., Licata, L.: *Detecting Outliers: Do Not Use Standard Deviation Around the Mean, Use Absolute Deviation Around the Median*. Journal of Experimental Social Psychology, Vol. 49, No. 4, pp. 764-766 (2013).
16. Knorr, E.M., Ng, R.T.: *Algorithms for Mining Distance-Based Outliers in Large Datasets*. VLDB-98, Proceedings of the 24rd Intern. Conference on Very Large Data Bases (1998).
17. Breunig, M.M., Kriegel, H., Ng, R.T., Sander, J.: *LOF: Identifying Density-Based Local Outliers*. Proceedings of the ACM SIGMOD Conference on Management of Data (2000).
18. Campos, G.O., Zimek, A., Sander, J., Campello, R.J.G.B., Micenková, B., Schubert, E., Assent, I., Houle, M.E. (2016). *On the Evaluation of Unsupervised Outlier Detection: Measures, Datasets, and an Empirical Study*. Data Mining and Knowledge Discovery, Vol. 30, No. 4, pp. 891-927.
19. Schölkopf, B., Williamson, R., Smola, A., Shawe-Taylor, J., Platt, J.: *Support Vector Method for Novelty Detection*. Advances in Neural Information Processing, Vol. 12, pp. 582-588 (2000).
20. Liu, F.T., Ting, K.M., Zhou, Z.: *Isolation-Based Anomaly Detection*. ACM Transactions on Knowledge Discovery from Data, Vol. 6, No. 1 (2012).
21. Shyu, M.L., Chen, S.C., Sarinnapakorn, K., Chang, L.W.: *A Novel Anomaly Detection Scheme Based on Principal Component Classifier*. Proceedings of the ICDM Foundation and New Direction of Data Mining workshop, pp. 172-179 (2003).
22. Pimentel, M.A.F., Clifton, D.A., Clifton, L., Tarassenko, L.: *A Review of Novelty Detection*. Signal Processing, Vol. 99, pp. 215-249 (2014).

23. Keogh, E., Lonardi, S., Ratanamahatana, C.A.: *Towards Parameter-Free Data Mining*. Proceedings of the Tenth ACM SIGKDD International Conference on Knowledge Discovery and Data Mining, Seattle, USA (2004).
24. Goldstein, M., Dengel, A.: *Histogram-based Outlier Score (HBOS): A fast Unsupervised Anomaly Detection Algorithm*. Proceedings of the 35th German Conference on Artificial Intelligence (KI-2012), pp. 59-63 (2012).
25. Foorthuis, R.: *SECODA: Segmentation- and Combination-Based Detection of Anomalies*. Proceedings of the 4th IEEE International Conference on Data Science and Advanced Analytics (DSAA 2017), Tokyo, Japan, pp. 755-764 (2017). DOI: 10.1109/DSAA.2017.35
26. Aggarwal, C. C., Yu, P.S.: *An Effective and Efficient Algorithm for High-Dimensional Outlier Detection*. The VLDB Journal, Vol. 14, No. 2, pp 211–221 (2005).
27. Dougherty, J., Kohavi, R., Sahami, M.: *Supervised and Unsupervised Discretization of Continuous Features*. Proceedings of the Twelfth International Conference on Machine Learning (1995).
28. Kotsiantis, S., Kanellopoulos, D.: *Discretization Techniques: A Recent Survey*. GESTS International Transactions on Computer Science and Engineering, Vol. 32, No. 1, pp. 47-58 (2006).
29. Foorthuis, R.: *Anomaly Detection with SECODA*. Poster Presentation at the 4th IEEE International Conference on Data Science and Advanced Analytics (DSAA), Tokyo (2017). DOI: 10.13140/RG.2.2.21212.08325
30. Yang, Y., Webb, G.I., Wu, X.: *Discretization Methods*. In: Maimon, O., Rockach, L. (Eds.), Data Mining and Knowledge Discovery Handbook. Kluwer Academic (2005).
31. Li, H., Hussain, F., Tan, C.M., Dash, M.: *Discretization: An Enabling Technique*. Data Mining and Knowledge Discovery, Vol. 6, No. 4, pp. 393–423 (2002).
32. Wolpert, D.H., Macready, W.G.: *No Free Lunch Theorems for Search*. Technical Report SFI-TR-95-02-010, Santa Fe Institute (1996).
33. Clarke, B., Fokoué, E., Zhang, H.H.: *Principles and Theory for Data Mining and Machine Learning*. Springer, New York (2009).
34. Rokach, L., Maimon, O.: *Data Mining With Decision Trees: Theory and Applications*. Second Edition. World Scientific Publishing, Singapore (2015).
35. Janssens, J.H.M.: *Outlier Selection and One-Class Classification*. PhD Thesis, Tilburg University (2013).
36. Maxion, R.A., Tan, K.M.C.: *Benchmarking Anomaly-Based Detection Systems*. International Conference on Dependable Systems & Networks, New York, USA (2000).
37. LAK, "Anomaly Detection at the Dutch Alliance on Income Data and Taxes". URL: www.loonaangifteketen.nl (2018).
38. Pijnenburg, M., Kowalczyk, W.: *Singular Outliers: Finding Common Observations with an Uncommon Feature*. Proceedings of the 17th International Conference on Information Processing and Management of Uncertainty in Knowledge-Based Systems (IPMU 2018), Cádiz, Spain; Springer CCIS 855 (2018).
39. Greenacre, M., Ayhan, H.: *Identifying Inliers*. Barcelona GSE Working Paper Series (2014).